# SiGe Raman spectra vs. local clustering/anticlustering : Percolation scheme and *ab initio* calculations

O. Pagès[1,*] R. Hajj Hussein[1] and V. J. B. Torres[2]

[1] *LCP-A2MC, Institut Jean Bariol, Université de Lorraine, Metz, 57078 Metz, France*
[2] *Departamento de Física and I3N, Universidade de Aveiro, Campus Santiago, 3810 – 193, Aveiro, Portugal*

**Abstract**

We formalize within the percolation scheme, that operates along the linear chain approximation, i.e., at one dimension (1D), an intrinsic ability behind Raman scattering to achieve a quantitative insight into local clustering/anticlustering in an alloy, using SiGe as a case study. For doing so, we derive general expressions of the individual fractions of the six SiGe percolation-type oscillators [1x(Ge-Ge), 3x(Si-Ge), 2x(Si-Si)], which monitor directly the Raman intensities, via a relevant order parameter $k$. This is introduced by adapting to the 1D-oscillators of the SiGe-diamond version of the 1D-percolation scheme, i.e. along a fully consistent 1D treatment, the approach originally used by Verleur and Barker for the three-dimensional (3D-)oscillators of their 1D-cluster scheme applying to zincblende alloys [H.W. Verleur and A.S. Barker, Phys. Rev. **149**, 715 (1966)], a somehow problematic one in fact, due to its 3D–1D ambivalence. Predictive $k$-dependent intensity-interplays between the $Si_{0.5}Ge_{0.5}$ Raman lines are confronted with existing experimental data and with *ab initio* Raman spectra obtained by using large (32-atom) disordered supercells matching the required $k$ values, with special attention to the Si-Ge triplet and to the Si-Si doublet, respectively.

[*] Author to whom correspondence should be addressed :
 olivier.pages@univ-lorraine.fr

**I. Introduction**

As soon as departing from pure (mono-constituent) media and considering mixtures, even involving only two substituent species, i.e. of the $A_{1-x}B_x$ type, one faces the key issue as how A and B arrange each other at a given composition x. Is the A-to-B substitution random, or is there any specific tendency for a given species (say A) to remain in its own local environment (A-like), or on the contrary to adopt a local environment mainly of the other (B) type, for some reason? Such deviations with respect to the ideal case of random A-to-B substitution are currently referred to as local clustering or local anticlustering, respectively.

Semiconductor mixed crystals, of the zincblende type, like $Zn_{1-x}Be_xSe$, or of the diamond type, with $Si_{1-x}Ge_x$ as leading system, can be considered as benchmark materials to address such issue.[1] Indeed they contain only two substituent species, each reducing to an atom (and not to a complicated molecule), moreover disposed on a quasi regular and isotropic (cubic) lattice, and attached to their immediate neighbors via strong (covalent) bonds.

In particular, the latter point concerning the chemical bonding is interesting with respect to local clustering/anticlustering (abbreviated c/ac hereafter). This is because the strength of a covalent bond, as currently measured at the laboratory scale via optical vibrational (phonon) spectroscopies, such as Raman scattering or infrared (IR) absorption, is highly sensitive to its local environment. A naïve rule is that *the bond force constant, and thus the phonon frequency,*[2] *falls down when a covalent bond is stretched, and vice versa*. For example, a spectacular Raman shift of ~50 cm$^{-1}$ is detected for the Be-Se bond in $Zn_{1-x}Be_xSe$, depending on whether a Be atom is isolated or paired to another Be atom (via an intermediary Se atom) in an otherwise pure ZnSe crystal (x~0).[3] In contrast, the corresponding difference in Be-Se bond length, estimated around 2% from *ab initio* calculations,[3] could not be resolved in recent extended x-ray absorption fine structure (EXAFS) measurements on a synchrotron.[4] This demonstrates that the bond force constant is potentially a more sensitive probe than the bond length to investigate the local environment of a bond. Apart from phonon spectroscopies (addressing the bond force constant) and EXAFS measurements (addressing the bond length), one may be tempted by x-ray diffraction (addressing the lattice constant).[1] However, the latter technique may be useful only when local c/ac leads to some periodical modulation of the alloy composition along certain crystal directions, corresponding to the formation of superlattice-like atom arrangements, at least on a restricted length scale. In this case, novel x-ray lines, reflecting the periodicity of the superlattice, are expected on top of the primitive ones due to the crystal lattice itself. However, not all deviations from random substitution may lead to the emergence of such long range ordering. This might rather be the exception than the rule. Therefore, in most cases x-ray diffraction might just be inappropriate. What remains then is our original suggestion to use the bond force constant. Thanks to its local character, it potentially constitutes a versatile probe, likely to address any type of local c/ac, in principle.

An interesting question is how the bond force constant can be used in practice to address local c/ac in an alloy? For simplicity, we start the discussion with $A_{1-x}B_xC$ zincblende alloys, in which c/ac preserves the A-C (1-x) and B-C (x) bond fractions. This is because the A and B substituting species remain bonded to the same intercalated C atom in a zincblende alloy, whether the A↔B substitution is random or not. In this case, one prerequisite to investigate local c/ac via Raman/IR spectroscopies is to have at one's disposal a distinct multi-mode signal per bond (explanation is given below). This makes sense in an alloy since the like bonds of a given species experience different distortions depending on their local environment, as needed to accommodate the local contrast in the bond physical properties (length, stiffness, ionicity…) of the coexisting A-C and B-C species. This may lead to as many Raman/IR lines per bond (recall the naïve rule quoted in italics above), and hence to a 1-bond→multi-mode pattern per bond, in principle. From there, provided a one-dimensional (1D) approach of the lattice dynamics of an alloy can be adopted, corresponding to a description of the alloy along the linear chain approximation (LCA), then one disposes of a univocal correspondence between a given Raman line/*frequency* and a given 'bond+environment' oscillator, to be discussed in terms of bond-stretching (1D). In this case, the individual fractions of such 1D-oscillators can be conveniently inferred from the *intensities* of the corresponding Raman lines, offering altogether, a detailed insight into the microstructure of an alloy. A basic conservation rule applies that the sum of the individual fractions of 1D-oscillators due to a given bond matches the corresponding bond fraction, which remains invariant in a zincblende alloy, as already mentioned.

Somewhat paradoxically, such procedure becomes irrelevant when using a more realistic three-dimensional (3D) description of the lattice dynamics. This is because the univocal correspondence between a given Raman line and a given 'bond+environment' motif, whereas valid at 1D, disappears at 3D. Indeed, a given 3D motif produces several Raman lines, corresponding to several variants of bond-stretching and bond-bending vibrations intra motif. Moreover, different 3D motifs may contribute to the same Raman line depending on the considered vibration patterns (refer e.g. to Fig. 3 in Ref. [5]). Altogether this makes it quasi impossible, in practice, to infer the fraction of a given 'bond+environment' 3D motif in the alloy from the intensities of the individual lines in the related multi-mode Raman/IR pattern, as extensively discussed, e.g., in Ref. [6].

Optical phonon spectroscopies, such as Raman scattering and infrared (IR) absorption are interesting regarding the possibility to adopt a 1D description of the lattice dynamics of an alloy. Indeed, due to the quasi vertical dispersion of the exciting light at the scale of the phonon Brillouin zone, the light-matter interaction takes place close to the Brillouin zone centre ($q \sim 0$). At this limit, the space-related phase term ($\vec{q} \cdot \vec{r}$) of a phonon disappears, and along with it an obligation to consider the actual position ($\vec{r}$) of an atom in the real 3D crystal. A scalar description of the lattice dynamics at 1D along the LCA suffices in principle. This has opened the way for three meaningful LCA-based (1D) models for the description of the Raman/IR of an alloy. Their main features are summarized below. We emphasize that Raman/IR spectroscopies detect only optic phonons at the Brillouin zone centre, corresponding to out-of-phase displacements of the intercalated *fcc* sublattices forming a zincblende structure or a diamond one, both taken as quasi rigid ones.

In fact the LCA-based models being used for decades to explain the Raman/IR spectra of random alloys, as originally developed for zincblende ones, either deny the existence of a 1-bond→multi-mode pattern for such systems, which eliminates the *sine qua non* condition to access local c/ac via vibrational spectra, as explained above; or, when such multi-mode pattern is actually formalized, the underlying approach is not free from conceptual ambiguity, and thus misleading *in fine* regarding the nature of the alloy disorder. Detail is given hereafter.

The modified-random-element-isodisplacement (MREI) model[7] comes under the first category. It assumes that, as soon as departing from the dilute limit of a *random* alloy, the original variety in the vibration pattern of a given impurity atom (recall the splitting between the Raman modes due to an isolated Be atom and to an isolated Be-pair fragment in an otherwise pure ZnSe crystal), disappears into a unique, possibly broadened, Raman/IR feature, corresponding to a basic 1-bond→1-mode (2-mode in total) pattern. The alternative *cluster* model,[8] worked out for those presumed *non-random* alloys that do obviously exhibit more than one phonon mode per bond in their Raman/IR spectra such as $Ga_{1-x}As_xP$, falls into the second category. It distinguishes between like bonds depending on their first-neighbor shell, out of four possible ones in a zincblende alloy, leading to a generic 1-bond→4-mode (8-mode in total) pattern. However, it seems difficult to conciliate an essentially–1D approach of the lattice dynamics within the *cluster* model – as testified by scalar equations of motion per atom – with an assignment of the elementary oscillators at 3D. In fact, fair contour modeling of the multi-mode IR spectra of $Ga_{1-x}As_xP$ could be achieved within the *cluster* model only by assuming a far-from-random As↔P substitution, while $Ga_{1-x}As_xP$ appears to be random by using other techniques. This has lead to challenge seriously the *cluster* model in recent years.[9]

In view of this, we may say that the presumed ability behind optical phonon spectroscopies to address local c/ac in a zincblende alloy remains unexploited in the traditional MREI and *cluster* schemes.

Over the past decade an alternative LCA-based (1D) model has been developed, the so-called *percolation* model. This has lead to a unified understanding within a generic 1-bond→2-mode (4-mode in total) behavior, of the long-standing MREI/*cluster*-based classification of the Raman/IR spectra of $A_{1-x}B_xC$ zincblende alloys including three main MREI sub-types (pure 2-mode, modified 2-mode and 1-mode), plus a generic multi-mode type covered by the *cluster* scheme.[5,10] In brief, the percolation model distinguishes between AC- and BC-like environments for a given A-C or B-C bond-stretching. What we retain is that this model combines the advantages of formalizing a proper multi-mode behavior per bond, as required to address local c/ac (see above) – in contrast with the MREI model, together with a consistent definition of the elementary oscillators at 1D, so as to remain in the spirit of the LCA – in contrast with the *cluster* model (see above).

It is worth to mention that recent inelastic neutron scattering measurements performed with the model percolation-type $Zn_{1-x}Be_xSe$ zincblende alloy have shown that the well-resolved Be-Se percolation doublet evidenced by Raman scattering at the Brillouin zone center (corresponding to quasi infinite phonon wavelengths), survives throughout the whole Brillouin zone up to the zone edge (corresponding to phonon wavelengths comparable with interatomic distances).[11] This establishes, on an experimental basis, that the percolation doublet finds its origin at the ultimate length scale of the very local environment of a bond, thus well-suited in principle to probe local c/ac in an alloy.

It has already been shown how the *percolation* scheme can be used, in practice, to reveal a trend towards local clustering in $Zn_{1-x}Be_xSe$ layers grown out of equilibrium conditions by using epitaxial techniques compared with nominally random single crystals of the same compositions grown from the melt under equilibrium conditions.[12] However, the discussion remained qualitative only. In particular, no order parameter could be introduced. This was due to an inability at the time to identify properly the microstructure of the *percolation*-type AC- and BC-like 1D-environments of a bond, in terms of both length scale and composition.

Recently, a decisive forward step has allowed us to overcome such major drawback at the occasion of a successful generalization of the *percolation* scheme beyond zincblende alloys, to the leading diamond-type $Si_{1-x}Ge_x$ alloy.[6]

Eventually the *percolation*-type oscillators of a $A_{1-x}B_xC$ zincblende alloy, in which, say, B-C refers to the short species, could be re-assigned in a fully explicit manner, according to $[(A-C)_1^B, (A-C)_1^A, (B-C)_1^B, (B-C)_1^A]$ – in which series the oscillators are ranked by order of increasing frequency, corresponding to $[C(A-C)B, C(A-C)A, C(B-C)B, C(B-C)A]$ in a developed 1D-notation. In the former (compact) notation, the main term refers to a given bond-stretching in a given 1D-environment whose length scale (first or second neighbors) and composition (AC- or BC-like) are specified via a subscript (1,2) and a superscript (A,B), respectively. Using the same (main term, subscript, superscript, ranking) code, the percolation scheme for $Si_{1-x}Ge_x$ consists of the following six basic oscillators $[\underline{Ge-Ge}, (Si-Ge)_2^{Ge}, \underline{(Si-Ge)_2^{SiGe}}, (Si-Ge)_2^{Si}, (Si-Si)_1^{Ge}, \underline{(Si-Si)_1^{Si+SiGe}}]$. The underlined modes correspond to the so-called main Ge-Ge, Si-Ge and Si-Si modes in a crude 1-bond→1-mode MREI-like description of the SiGe Raman pattern. The alternative (not underlined) modes remain minor over most of the composition domain in the random $Si_{1-x}Ge_x$ alloy,[6] and are currently referred to as the fine structure of the SiGe Raman spectra.

We can be more explicit, for future use, by adding that the lower and upper Si-Ge oscillators are due to all-Ge (lower branch) and all-Si (upper branch) environments, while the remaining (Si,Ge)-mixed environments give rise to a common oscillator in between (intermediary branch). We mention that in the random SiGe alloy the upper Si-Ge mode decomposes into a multiplet as soon as departing from the Si-parent limit, say beyond 30 at.% Ge. The assignment of such multiplet is not clear yet, and a persisting source of problem when comparing experimental/*ab initio* Si-Ge Raman data with calculated percolation-type Raman lineshapes, as done e.g. in **Sec. IV** of the present work.

The Si-Si stretching, as for it, distinguishes between all-Ge (bottom branch) and alternative Si-like environments (top branch), i.e. including at least one Si atom. As for Ge-Ge, its stretching produces a unique Raman line at any alloy composition, reflecting a basic insensitivity of the Ge-Ge stretching to its local environment. Such detailed assignment of the percolation-type environments of a bond has allowed us to derive fully explicit fractions of the individual percolation-type 1D-oscillators depending on the alloy composition x for *random* $Si_{1-x}Ge_x$. **Fig. 1** summarizes the content of the percolation scheme per oscillator for random SiGe, regarding both the Raman-intensity, as monitored by the corresponding fraction of individual oscillator (general expressions explained in **Sec. III-1** are reported in the body of the figure – see the $f_i$ terms, where $i$ refers to a given oscillator/branch as specified by black symbols), and the Raman-frequency (plain lines). Extensive detail is available in Ref. **6**.

Our ambition in this work is to achieve a versatile version of the percolation scheme for the lattice dynamics of an alloy, in which the individual fractions of oscillators, which monitor directly the Raman intensities, take a general form depending on a relevant parameter $k$ of local c/ac, that remains to be defined. We emphasize that, by construction, the present approach is not relevant for the discussion of any c/ac-induced change in the Raman frequencies; only the Raman-intensity aspect is covered.

For working out such generalized $k$-dependent version of the percolation scheme, we focus on the diamond-type $Si_{1-x}Ge_x$ alloy, and not on a $A_{1-x}B_xC$ zincblende alloy. Indeed we anticipate that the task may be more demanding, and thus at the same time more general, for SiGe, corresponding to three bond species (Ge-Ge, Si-Ge, Si-Si), and six percolation oscillators in total, with bond environments defined up to second neighbors,[6] than for any zincblende alloy, in which two bond species (A-C and B-C) offer only three percolation oscillators, moreover restricted to the first-neighbor environment of a bond (see Refs. **5** and **10**). Additional complication may arise in that, in a diamond alloy such as $Si_{1-x}Ge_x$, the Ge-Ge, Si-Ge and Si-Si bond fractions, which respectively scale as $x^2$, $2x \cdot (1-x)$ and $(1-x)^2$ in the random alloy, modify when c/ac develops. This is because any site of the diamond lattice is likely to be occupied by Si or Ge atoms, which thus may 'see each other' as immediate neighbors, with concomitant impact on the Si-Si, Si-Ge and Ge-Ge bond fractions, and thus also on the corresponding Raman intensities. For example, when assuming a random substitution at x=0.5, the Ge-Ge, Si-Ge and Si-Si bond fractions amount to 0.25, 0.5 and 0.25, respectively, and the Si-Ge mixed-bond dominates. In case of full clustering, the Si-Ge bond just disappears – if we neglect the minor fraction at the frontier between the phase-separated Si and Ge semi-infinite crystals. In contrast, local c/ac preserves the A-C and B-C bond fractions in a $A_{1-x}B_xC$ zincblende alloy, as already mentioned. In terms of oscillator strength, which physical parameter varies linearly with the fraction of oscillator,[7] we may say that, in a diamond alloy such as SiGe, inter-bond transfer of oscillator strength, i.e. taking place in between Raman modes due to distinct bond species, superimposes onto the zincblende-like intra-bond transfer of oscillator strength driven by local c/ac, as merely concerned with the like Raman modes due to a given bond. In view of all this, the case of SiGe can be identified as a sort of bottleneck with respect to our ambition in this work to formalize the presumed sensitivity of optical phonon spectroscopies to local c/ac in an alloy. We are confident that, provided such generalized $k$-dependent version of the percolation scheme can be worked out for the diamond-type SiGe alloy, then its transposition to any zincblende alloy should be straightforward.

In practice, the parameter $k$ of local c/ac is introduced in our SiGe percolation scheme following the approach earlier used by Verleur and Barker in their 1D *cluster* model that applies to the first neighbor 3D-environments of a bond in a zincblende alloy (see above).[8] In brief, such parameter $k$ monitors the probability of finding a given atom besides another atom of the like species. Basically the probability is larger than the related fraction of atoms in the alloy in case of clustering ($k>0$), and smaller in case of anticlustering ($k<0$). Extensive detail is given in **Sec. III-1**. Now, as already mentioned, the approach of Verleur and Barker is not free from conceptual ambiguity, due to its 1D-3D ambivalence. In order to remove such ambiguity, we opt for a fully consistent 1D approach, for sake of consistency with the LCA, upon which the percolation scheme relies. Moreover the approach is presently generalized to a diamond alloy, i.e. SiGe, and extended to second-neighbor environments besides first-neighbor ones, in reference to the Si-Ge triplet and to the Si-Si doublet, respectively.

The manuscript is organized as follows. Based on a selected experimental data in the literature, we identify in **Sec. II** the main features characterizing the dependence of the SiGe Raman spectra on c/ac, in view to test the predictions of the $k$-dependent version of the SiGe percolation scheme. In **Sec. III**, we derive such generalized version of the SiGe percolation scheme, in which the individual fractions of 1D-oscillators are expressed via a proper parameter $k$ of local c/ac. This is introduced along the approach used by Verleur and Barker in their 1D *cluster* model that applies to the first neighbor 3D-environments of a bond in a zincblende alloy.[8] Only, in order to remove an ambiguity behind such 1D-3D ambivalence, we adopt a pure 1D formalism in this work, as detailed in **Sec. III-1**. This applies to the diamond structure, using $Si_{1-x}Ge_x$ as a case study, and, further, extends beyond the first-neighbor 1D-environment of a bond, up to the second-neighbor one, in reference to the Si-Si doublet and to the Si-Ge triplet, respectively. In **Sec. III-2** predictive intensity-interplays between the $Si_{0.5}Ge_{0.5}$ Raman lines depending on clustering ($k>0$) and anticlustering ($k<0$) are produced both within the crude MREI-like scheme – thus only concerned with the three main Ge-Ge, Si-Ge and Si-Si lines, for reference purpose, and also within the more refined percolation scheme – covering the main Raman features as well as the Si-Ge and Si-Si fine structures, corresponding to six oscillators in total [1×(Ge-Ge), 3×(Ge-Ge), 2×(Si-Si)].[6] The resulting $k$-dependent percolation-type lineshapes are eventually confronted with part of the earlier selected experimental data (refer to **Sec. II**) on the one hand, in **Sec. III-3**, mostly in reference to the Si-Ge triplet then, and with corresponding *ab initio* Raman spectra calculated with the AIMPRO code by using reasonably large (32-atom) fully-relaxed supercells matching the required $k$ values, on the other hand, in **Sec. IV**, searching then for a special insight into the Si-Si doublet. A prerequisite for the discussion of the *ab initio* spectra is a proper assignment of the Ge-Ge, Si-Ge and Si-Si spectral ranges depending on $k$. This is achieved in **Sec. IV-1** based on *ab initio* calculation of limit Raman frequencies arising from prototype impurity motifs in case of full c/ac. The effect of c/ac on the *ab initio* Raman intensities is then discussed in **Sec. IV-2**, notably in its adequacy with the prediction of the percolation scheme. Last, a brief discussion of the *ab initio* Raman frequencies in their $k$-dependence is also included, in **IV-3**, for sake of completeness. Conclusions are summarized in **Sec. V**.

**II. SiGe Raman spectra in their dependence on clustering/anticlustering – Main features in the literature**

When searching for Raman data in the literature to be used as references in this work, we observe that basically all the existing experimental[13-23] and theoretical[21-26] studies of local c/ac so far reported with $Si_{1-x}Ge_x$ take place around x=0.5, within 10%. Recently, the effect of c/ac was explored with alternative SiGe alloy compositions,[27] but mostly in relation to the Raman frequencies, while our main concern in this work is the Raman intensities, as already explained. Interestingly, the three bond species are in comparable proportions in the random $Si_{0.5}Ge_{0.5}$ alloy (refer above). In fact, the three main Ge-Ge, Si-Ge and Si-Si Raman lines (underlined in the percolation series) exhibit similar intensities. Moreover, the three minor modes constituting the fine structure (not underlined), though comparatively weak, are clearly visible as well (for an overview see Fig. 5 of Ref. **6**). At any other x value, at least one of the minor mode is missing. Therefore, x~0.5 is ideal when searching to detect any possible c/ac-induced deviation with respect to the nominal intensity balance between the SiGe Raman lines as observed in a random alloy. Additional interest arises in that x=0.5 is the only alloy composition at which both clustering and anticlustering can fully develop (detail is given in **Sec. III-2**).

On the theoretical side, all types of atom arrangements can be generated in a supercell approach, in principle, provided the used supercell is large enough. This is not as simple experimentally. Intuitively, one may start to grow either a $(Si)_n/(Ge)_n$ superlattice, a short-period one so as to ensure a certain proximity of the Si and Ge atoms, or a zincblende SiGe compound, corresponding to limit cases of full clustering and full anticlustering,[28] respectively. Then, progressive interdiffusion of Si and Ge might be finely tuned by applying some relevant cycle of thermal annealing until, at a certain stage, a random (Si,Ge) distribution is eventually achieved. We may already forget about the anticlustering→random route, simply because the starting zincblende SiGe crystal does not exist in reality. In contrast, high-quality short-period $Si_nGe_n$ superlattices can be grown since the mid-eighties by using epitaxial techniques.[14–23,27] This has offered a possibility to explore the alternative clustering→random route in several occasions, as summarized below.

We distinguish two different regimes along the clustering→random route, i.e. a 'superlattice regime' in the early stages of the annealing process, in which a certain long range order is somehow preserved, and a 'disordered regime' after intense annealing, in which the long range order is destroyed, resulting in a basically disordered alloy, not exclusive, though, of a reminiscent trend towards local clustering. The key point is that the two regimes correspond to different Raman patterns. The 'superlattice regime', abundantly studied in the literature, both experimentally[17–23] and theoretically,[21–26] is characterized by specific Raman features, such as confined optical modes in the Si- or Ge-like layers, accompanied by acoustical modes resulting from zone folding due to the additional periodicity of the superlattice.[19–23] Such complex vibration pattern falls beyond the scope of the percolation scheme, and is not addressed in this work.

We focus on the 'disordered regime', which starts as soon as the Raman pattern basically fits into the six-oscillator percolation scheme, and resembles that of a random $Si_{0.5}Ge_{0.5}$ alloy. Short-period superlattices with smeared, and not sharp, Si/Ge interfaces, reflecting significant Si↔Ge intermixing, enter this category. We are aware of two detailed Raman studies on such systems, both concerned with $Si_4Ge_4$ superlattices.[20,21] Short-period composition-modulated alloys – simply referred to as modulated alloys hereafter, corresponding to alternation of SiGe bi-layers with alloy compositions complementary to 0.5 into a superlattice sequence, such as those envisaged by Tsang et al. in their careful Raman study,[16] are even better candidates for our use. This is because each bi-layer already consists of a disordered alloy (detail is given in **Sec. III-3**), and not of a nominally pure Si or Ge layer as in the previous two examples of $Si_4Ge_4$ superlattices. Altogether, the experimental Raman spectra taken by Schorer et al.,[20] Alonso et al.[21] and Tsang et al.[16] constitute the corpus of $Si_{0.5}Ge_{0.5}$-related data that we will use to test the prediction of the generalized $k$-dependent version of the SiGe percolation scheme. The main trends observed with these systems are the following.

When clustering reduces, the main Si-Ge mode reinforces at the cost of the Ge-Ge and main Si-Si ones (underlined in the percolation series, see above), as expected within a crude MREI-like description. This simply reflects the reinforcement/disappearance of the corresponding bonds along the 'clustering→random' route, respectively. From the magnitude of the deviation between the observed intensity ratios and the nominal ones for a random alloy, we may infer that the clustering is more pronounced for the $Si_4Ge_4$ superlattices (large deviation) than for the modulated alloy (small deviation). Nevertheless, in the latter case the trend shows up unambiguously in the point-by-point difference spectrum carefully elaborated by Tsang et al. by subtracting to the Raman spectrum of their as-grown bilayer $Si_{0.5}Ge_{0.5}$ duperlattice, corresponding to moderate clustering, the one obtained with the same sample but after destruction of the long range order via intense annealing, thus taken as nominally random.[16] For a direct insight, the difference (thick curve) and reference (thin curve, shifted beneath) spectra in question are digitalized and reproduced in **Fig. 2a**.

We want to be more accurate and turn now to the Si-Ge and Si-Si fine structures (not underlined in the percolation series, see above), whose understanding falls beyond the MREI scheme. The dependence of such fine structure on c/ac has attracted little attention so far.

In the Si-Ge triplet, we observe that residual shoulders emerge on each side of the main/central Si-Ge mode in the difference Raman spectrum elaborated by Tsang et al. (refer to the thick arrows in **Fig. 2a**). This indicates that the collapse of the latter mode with clustering (see above) goes with reinforcement of its two minor satellites.

As for the Si-Si doublet, careful examination of the difference Raman spectrum of Tsang *et al.*, does not reveal any significant change in the intensity of the lower/minor Si-Si satellite (refer to the dashed line in **Fig. 2a**) while the upper/main Si-Si mode strengthens (see above). The comparison between the Raman spectra of the highly-ordered $Si_4Ge_4$ superlattices and of a reference random alloy (refer to the two bottom spectra in Fig. 1 of Ref. **19**), as reproduced in **Fig. 2b**, is more fruitful with this respect. It is striking that the minor Si-Si feature, which shows up as a distinct shoulder besides the main Si-Si mode in the Raman spectrum of the reference random alloy (bottom spectrum), has totally disappeared with the $Si_4Ge_4$ clustered ones (upper spectrum, refer to the arrow). This indicates that, under clustering, the reinforcement of the upper/main Si-Si mode (see above) is accompanied by a progressive disappearance of its lower/minor Si-Si satellite.

Summarizing, our selected data reveal the following intensity-interplays between the SiGe Raman lines when clustering develops. First, **(i)** the unique Ge-Ge and upper/main Si-Si features reinforce at the cost of the central/main Si-Ge one. This goes with **(ii)** reinforcement of the satellite Si-Ge features, and **(iii)** progressive disappearance of the lower/minor Si-Si satellite. By symmetry, we anticipate opposite trends in case of anticlustering, until in the final stage, corresponding to the formation of a pure zincblende SiGe crystal, i.e. the limit case of anticlustering, only the central/main Si-Ge mode survives. The $k$-dependent version of the SiGe percolation scheme that we aim at deriving in this work should naturally incorporate features **(i)**–**(ii)**–**(iii)**. This is a criterion for validation.

### III. Generalized $k$-dependent version of the SiGe percolation scheme

C/ac effects are taken into account in the diamond-SiGe version of the percolation scheme by adapting to the corresponding 1D oscillators, as defined at the bond scale (Ge-Ge) or at the larger scale of the first (Si-Si) and second (Si-Ge) neighbors of a bond, the approach originally developed by Verleur and Barker in their 1D *cluster* model for the 3D oscillators referring to all possible first neighbor environment of a bond in a zincblende alloy (four in total).[8] A basic difference with their approach, though, is that we opt for a pure 1D formalism, for sake of consistency with the LCA upon which the percolation scheme relies.[6]

1. **Expression of the individual fractions of percolation-type SiGe 1D-oscillators in their dependence on $k$**

A relevant parameter $k$ of local c/ac is introduced by materializing a trend towards c/ac of a given substituting species in $Si_{1-x}Ge_x$, the trend being emphasized when the species gets more and more dilute. For example, we define $P_{SiSi}$ as the probability of finding one Si atom next to another Si atom on the (Si,Ge)-diamond lattice at 1D. In the random alloy $P_{SiSi}$ is not dependent on the local neighborhood of an atom, thus equal to the probability $P_{Si}$ of having one Si atom on a given site, which identifies with the fraction $(1 - x)$ of Si atoms in the crystal. The trend towards local c/ac corresponds to an increased/decreased probability $P_{SiSi}$ with respect to the random case when the fraction of Ge atoms ($x$) increases. Accordingly $P_{SiSi}$ may simply write as

$$P_{SiSi} = (1 - x) + k_{1,Si} \cdot x. \tag{1a}$$

where the first subscript of $k$ refers to the number of atoms added by the 'starting' one to form the uniform (Si-like) 1D cluster under consideration (in this case, only one atom is added), and the second subscript refer to the cluster type (Si- or Ge-like). Depending on the sign of $k_{1,Si}$ this expression may as well be used to describe clustering ($k_{1,Si} > 0$) or anti-clustering ($k_{1,Si} < 0$, with some limitation then, as discussed below). A similar expression can be likewise defined for $P_{GeGe}$, corresponding to

$$P_{GeGe} = x + k_{1,Ge} \cdot (1 - x). \tag{1b}$$

With a similar notation. As a tendency towards c/ac of a given minor substituting species necessarily accompanies the same tendency for the other (dominant) substituting species, we may infer that $k_{1,Si}$ and $k_{1,Ge}$ are related. The relation is expressed by equaling the two possible expressions of the probability of finding one Si atom next to one Ge atom in the crystal, with either Si or Ge as the 'starting' site. This writes as

$$P_{Si} \cdot P_{SiGe} = P_{Ge} \cdot P_{GeSi} \qquad (2)$$

equivalent to

$$(1-x) \cdot (1 - P_{SiSi}) = x \cdot (1 - P_{GeGe}), \qquad (3)$$

eventually leading to

$$k_{1,Si} = k_{1,Ge}, \qquad (4)$$

Abbreviated $k_1 = k$ hereafter.

We may likewise define $P_{(SiSi...Si)_n}$ and $P_{(GeGe...Ge)_n}$ for 1D clusters constituted of $n$ like Si or Ge atoms, as

$$P_{(SiSi...Si)_n} = P_{Si} + k_{n-1,Si} \cdot x \qquad (5a)$$
$$P_{(GeGe...Ge)_n} = P_{Ge} + k_{n-1,Ge} \cdot (1-x) \qquad (5b)$$

with the same convention for the subscripts of $k$. General forms of the $k_{n-1,Si}$ values for $n$ equal to 2 and 3, which appears to be sufficient for our purpose to derive generalized $k$-dependent individual fractions for the whole set of six percolation-type SiGe 1D–oscillators, may then be inferred from the following series of correspondences, similar in nature to Eq. (3):

$$P_{Si} \cdot P_{SiSi} \cdot P_{SiSiGe} = P_{Ge} \cdot P_{GeSi} \cdot P_{GeSiSi} \qquad \text{(access to } k_{2,Si}\text{)} \qquad (6a)$$
$$P_{Si} \cdot P_{SiSi} \cdot P_{SiSiSi} \cdot P_{SiSiSiGe} = P_{Ge} \cdot P_{GeSi} \cdot P_{GeSiSi} \cdot P_{GeSiSiSi} \qquad \text{(access to } k_{3,Si}\text{)} \qquad (6b)$$

In expressing such correspondences we assume, like in Ref. **8**, that

$$\forall X = (Si, Ge) \text{ and } \forall n, P_{GeSi(XX...X)_n} = P_{SiGe(XX...X)_n} = P_{(XX...X)_n} \qquad (7)$$

In fact Eq. (7) reflects a technical inability behind our specific way to introduce $k_{n,Si}$ to express any probability related to the formation of a given 1D cluster 'starting' by a non uniform succession of atoms. For example $P_{GeSiSi} = 1 - P_{GeSiGe}$ cannot be directly expressed nor via $P_{SiSiSi}$ neither via $P_{GeGeGe}$, but advantageously simplifies to $P_{Si} = 1 - x$ when applying (7), in consideration of what $P_{SiSiGe} = 1 - P_{SiSiSi}$ may be derived, giving access to $k_{2,Si}$. Symmetrical equations to Eqs. (6), in which Si is replaced by Ge and *vice versa*, are correspondingly used to access the $k_{n,Ge}$ values.

Using Eqs. (6) and (7), the following series of important probabilities may then be express via $P_{SiSi}$ and $P_{GeGe}$, whose dependencies on the order parameter $k$ are specified according to Eqs. (1),

$$P_{SiSiSi} = [P_{SiSi} - (1-x) \cdot (1 - P_{SiSi})] \cdot P_{SiSi}^{-1} \qquad (8a)$$
$$P_{SiSiSiSi} = [P_{SiSi} - (1-x) \cdot (1 - P_{SiSi}^2)] \cdot [P_{SiSi} - (1-x) \cdot (1 - P_{SiSi})]^{-1} \qquad (8b)$$

with similar expressions for $P_{GeGeGe}$ and $P_{GeGeGeGe}$ as obtained just by substituting Ge for Si in each subscript of Eq. (8). Such probabilities are the basic ingredients in the fractions of the individual 1D-oscillators of the SiGe percolation scheme. For example, the fraction of Si-Si 1D-oscillator corresponding to Si-Si stretching in a (Si,Ge)-mixed first-neighbor environment, labeled $(Si - Si)_1^{SiGe}$ in the terminology of the percolation scheme, expresses according to $f\{(Si - Si)_1^{SiGe}\} = 2P_{Si}P_{SiSi}P_{SiSiSi}P_{SiSiSiGe}$, where 2 refers to the number of possible orientations of the 1D-cluster under consideration, and $P_{SiSiSiGe} = 1 - P_{SiSiSiSi}$. We recall that the latter oscillator contributes, together with $(Si - Si)_1^{Si}$ that refers to Si-Si stretching in a pure-Si first-neighbor environment, to the upper branch of the (Si-Si) doublet in the SiGe percolation scheme. Finally, we obtain the following $k$-dependent expressions for the individual fractions of the six 1D-oscillators of the SiGe percolation scheme, hereafter ranked in order of increasing frequency:

$$f_1 = f\{(Ge - Ge)\} = x \cdot P_{GeGe} \qquad (9a)$$
$$f_2 = f\{(Si - Ge)_2^{Ge}\} = 2x^2 \cdot (1 - P_{GeGe}) \cdot [P_{GeGe} - x \cdot (1 - P_{GeGe}^2)] \qquad (9b)$$
$$f_3 = f\{(Si - Ge)_2^{SiGe}\} = 2(1-x) \cdot (1 - P_{SiSi}) - [f\{(Si - Ge)_2^{Ge}\} + f\{(Si - Ge)_2^{Si}\}] \qquad (9c)$$
$$f_4 = f\{(Si - Ge)_2^{Si}\} = 2(1-x)^2 \cdot (1 - P_{SiSi}) \cdot [P_{SiSi} - (1-x) \cdot (1 - P_{SiSi}^2)] \qquad (9d)$$
$$f_5 = f\{(Si - Si)_1^{Ge}\} = (1-x)^2 \cdot (1 - P_{SiSi})^2 \qquad (9e)$$
$$f_6 = f\{(Si - Si)_1^{SiGe+Si}\} = 2(1-x)^2 \cdot P_{SiSi} \cdot (1 - P_{SiSi}) + (1-x) \cdot [P_{SiSi} - (1-x) \cdot P_{SiSi}^2] \qquad (9f)$$

Based on such refined six-oscillator [1x(Ge-Ge), 3x(Si-Ge), 2x(Si-Si)] percolation-type description, one may easily infer a more crude three-oscillator [1x(Ge-Ge), 1x(Si-Ge), 1x(Si-Si)] MREI-like description. This comes to a vision in which the individual percolation-type three-mode and two-mode patterns of the Si-Ge and Si-Si bonds, respectively, virtually condense into an unique feature per bond. The $k$-dependent fractions of such presumed 1-bond→1-mode Ge-Ge, Si-Ge and Si-Si MREI-like 1D-oscillators would accordingly identify with

$$f\{(Ge-Ge)\} = f_1 \tag{10a}$$
$$f\{(Si-Ge)\} = P_{Si} \cdot P_{SiGe} + P_{Ge} \cdot P_{GeSi} = f_2 + f_3 + f_4 \tag{10b}$$
$$f\{(Si-Si)\} = P_{Si} \cdot P_{SiSi} = f_5 + f_6 \tag{10c}$$

Such expressions simply reflect basic conservation rules as concerned with the total fractions of Si-Ge and Si-Si bonds in the alloy for a given value of the order parameter $k$. Note that in contrast with the A-C (1-x) and B-C (x) bond fractions in a $A_{1-x}B_xC$ zincblende alloy, that remain stable under c/ac (as explained in **Sec. I**), the Ge-Ge, Si-Ge and Si-Si bond fractions of the $Si_{1-x}Ge_x$ diamond alloy, defined via Eqs. (10) are all $k$-dependent. Such equations incorporate both intra-bond and inter-bond transfers of oscillator strength in the diamond SiGe alloy, while only intra-bond transfer of oscillator strength exists in a zincblende alloy.

### 2. Percolation/MREI–like $k$-dependent $Si_{0.5}Ge_{0.5}$ Raman lineshapes

$k$-dependent Raman lineshapes for the representative $Si_{0.5}Ge_{0.5}$ alloy corresponding to intermediary composition are calculated along the exact procedure detailed in Ref. **6**, using the same input parameters, except that the individual fractions of oscillators $f_i$, which monitor the Raman intensities, are presently re-expressed by using the more general $k$-dependent forms given in Eqs. (9). In particular the individual Raman frequencies and linewidths are taken identical to the values found for the *random* $Si_{0.5}Ge_{0.5}$ alloy, simply because our formalism does not naturally incorporate any $k$-dependence neither for the Raman frequency, as already discussed in **Sec. I**, nor for the phonon damping, which is sample-dependent. Both clustering ($0 < k < 1$) and anticlustering ($0 > k > -1$) are considered besides the random case ($k = 0$, the corresponding fractions of oscillators are explicit, e.g., in Fig. 4 of Ref. **6**). Two distinct sets of curves in **Fig. 3** are displayed: one obtained within the six-oscillator SiGe percolation scheme (thick lines) using Eqs. (9), and one obtained within the more crude three-oscillator MREI-like scheme (thin lines) using Eqs. (10), added for comparison.

We emphasize that similar sets of curves can be derived at any alloy composition, in principle, not only at x=0.5. However, $Si_{0.5}Ge_{0.5}$ is particularly interesting in that the calculations can be pushed to the limit $k$ values, corresponding to full clustering ($k = +1$) and full anticlustering ($k = -1$). At any other alloy composition, anticlustering cannot fully develop. This is because when departing from x=0.5 the major bond species has to remain segregated to a certain extent, at least locally. Technically, the anticlustering limit is achieved when the calculated fraction of the minor bond species becomes negative when using Eqs. (9)-(10), which is not realistic physically. Such situation is never encountered in case of clustering, which can fully develop at any alloy composition.

The basic trend with clustering is that the Ge-Ge and Si-Si features reinforce at the cost of the intermediary Si-Ge one, consistently with experimental findings [refer to **(i)** in **Sec. II**]. Ultimately ($k = +1$), only the Ge-Ge and Si-Si signals survive. This reflects directly the abundance of the corresponding bonds in the crystal (see **Sec. I**). Careful examination of the fine structure in each spectral range further reveals that the Si-Ge side features reinforce at the cost of the central one (compare the MREI-thin and percolation-thick Si-Ge lines), until at a certain stage ($k \gtrsim 0.5$) both side features become dominant. This conforms to intuition since the side and central modes refer to pure (Ge or Si) and (Si,Ge)-mixed environments (see **Sec. I**), thus favored and disfavored by clustering, respectively. Again, this is consistent with experimental findings [refer to **(ii)** in **Sec. II**]. The trend is just opposite for the Si-Si doublet. The lower/minor Si-Si satellite weakens with clustering, to the benefit of the upper/main Si-Si feature that reinforces (compare the MREI-thin and percolation-thick Si-Si lines). This makes perfect sense when realizing that the latter minor and dominant features are due to Si-Si stretching in pure-Ge and Si-like environments, respectively (see **Sec. I**). The corresponding 'bond+1D-environments' oscillators, of the (Si,Ge)-mixed and Si-like types, are thus disfavored and favored by clustering, respectively. Again, this is consistent with experimental findings [refer to **(iii)** in **Sec. II**].

The $k$-dependence of the multi-mode SiGe Raman intensities in case of anticlustering can be understood along the same lines, leading to reverse trends for the main as well as minor modes. We do not find it useful to enter more detail.

### 3. Percolation vs. experimental insights – main (Ge-Ge, Si-Ge, Si-Si) features and (Si-Ge) triplet

In the following we briefly test the $k$-dependent version of the SiGe percolation scheme, in relation to points **(i) – (iii)** identified in **Sec. II**, on a quantitative basis. A natural experimental reference with this respect is the point-by-point difference Raman spectrum elaborated by Tsang *et al.*[16] (refer to the upper curve in **Fig. 2**) by substracting the Raman spectrum from their modulated $Si_{0.5}Ge_{0.5}$ bilayer alloy, corresponding to moderate clustering, from that of a reference $Si_{0.5}Ge_{0.5}$ random alloy (corresponding to the bottom curve in **Fig. 2**).

In view to elaborate a theoretical percolation-type difference Raman spectrum to compare with such experimental reference, we need to determine the relevant value of our vibrational-like order parameter $k$, as defined within the percolation scheme, corresponding to the structural-like order parameter $\eta$=0.64, as independently determined by Tsang *et al.* for their modulated $Si_{0.5}Ge_{0.5}$ alloy by using x-ray diffraction.[16] In fact this alloy consists of a superlattice-like periodical modulation of the alloy composition along the [111] crystal axis under the form of a $Ge_{0.5(1+\eta)}Si_{0.5(1-\eta)} - Ge_{0.5(1+\eta)}Si_{0.5(1-\eta)} / Ge_{0.5(1-\eta)}Si_{0.5(1+\eta)} - Ge_{0.5(1-\eta)}Si_{0.5(1+\eta)}$ bilayer alternation, in which closely (widely) spaced (111) planes are of the like (different) species. A simple bond-counting procedure leads to the following estimates of the individual bond fractions depending on $\eta$,

$f\{(Si - Si)\} = \frac{1}{8}\left\{0.5(1 + \eta) \cdot \left[0.5(1 + \eta) + \sum_{u=0}^{3} u \cdot \frac{3!}{u!(3-u)!} \cdot [0.5(1 - \eta)]^u \cdot [0.5(1 + \eta)]^{3-u}\right] + A\right\}$ (11a)

being clear that

$f\{(Ge - Ge)\} = f\{(Si - Si)\}$ (11b)

which implies

$f\{(Si - Ge)\} = 1 - 2f\{(Si - Si)\},$ (11c)

where $A$ represents a variant of the entire first term in Eq. (11a) in which $(1 - \eta)$ is replaced by $(1 + \eta)$, and *vice versa*. Accordingly, for $\eta$=0.64, the Ge-Ge, Si-Ge and Si-Si bond fractions amount to ~30%, ~40% and ~30%, respectively, to compare with 25%, 50% and 25% in the random alloy ($\eta$=0, see **Sec. I**). A similar distribution of bond fractions is achieved via Eqs. (10) by taking $k$~0.20.

We expect that the theoretical 'clustered-random' difference spectrum obtained by subtracting the percolation-type Raman spectrum due to the $Si_{0.5}Ge_{0.5}$ random alloy, defined by $k$~0, from the corresponding spectrum due to the partially-ordered alloy, characterized by $k$~0.20, more or less matches the experimental difference spectrum obtained by Tsang *et al*. For a direct comparison we superimpose in **Fig. 2a** the theoretical percolation-type Raman lineshapes obtained for the difference spectrum (top curve), and also for the reference random alloy (bottom curve), onto the corresponding experimental data of Tsang *et al*. In the latter case, the superimposition is done after normalization to the intensity of the unique Ge-Ge mode. Note that the theoretical percolation-type Raman lineshape for random $Si_{0.5}Ge_{0.5}$ is similar in every respect (phonon dampings, Raman frequencies and Raman intensities) to the corresponding generic curve earlier derived in Ref. **6** (refer to x=0.5 in Fig. 5 therein).

First we compare briefly the experimental and theoretical Raman spectra of the reference random alloy (refer to the bottom curves in **Fig. 2a**). We are forced to admit that there is a problem with the intensities of the main Si-Ge and Si-Si modes, the latter showing up much more strongly in the experimental spectrum than in the theoretical one. In fact, this must not be so surprising since Tsang *et al*. did record their reference Raman spectrum under near resonant condition, i.e. by using the 457.9 nm excitation at 300K, leading to significant enhancement of the Si-Si and Si-Ge Raman signals (refer to Figs. 2 and 3 of Ref. **20**) with respect to the Ge-Ge one, that remains basically insensitive to this particular resonance (Ref. **20**). Obviously, such resonance effects were not incorporated into the percolation scheme. We recall that the intrinsic Raman efficiencies of the Ge-Ge, Si-Ge and Si-Si bonds were directly calibrated by implementing *ab initio* calculations in pure diamond-Si, zincblende-SiGe and diamond-Ge supercells out-of-resonance condition, respectively.[6] In fact, the 'starting discrepancy' between experience and theory concerning the Raman spectrum of the random alloy is not so dramatic when realizing that our main purpose is to compare the experimental and theoretical 'clustered-random' difference Raman spectra (in reference to the top curves in **Fig. 2a**). In principle, the 'starting discrepancy' should be suppressed in the difference spectra.

In fact, the experimental and theoretical 'clustered-random' difference spectra appear to be remarkably consistent in view of the simplicity of our percolation approach, especially in what concerns the intensities of the main features and of the Si-Ge triplet. The basic experimental trend, i.e. a reinforcement of the unique Ge-Ge and upper/main Si-Si modes, accompanied by a collapse of the main/central Si-Ge mode, is well-reproduced in the percolation-like theoretical curve. More interestingly, the two small resonances that emerge on each side of the main Si-Ge antiresonance in the experimental difference show up clearly in the percolation-type difference. The situation is not as clear with the Si-Si doublet. Indeed the theoretical antiresonance due to the lower/minor Si-Si mode does not show up in the experimental curve. Our present view is that the $k$-difference between the modulated alloy ($k$~0.2) and the random one ($k$~0), is not sufficiently large to generate unambiguous impact on the Raman intensity of the latter mode, a rather broad/overdamped one in fact. Additional complication arises in that the main Si-Si Raman signal exhibits a significant blue-shift with clustering (see the double horizontal arrow in **Fig. 2**), which cannot be accounted for within our percolation approach, as already explained. While the shift in question is rather small, i.e. of the order of 5 cm$^{-1}$, it is not negligible for all that, representing approximately one-third of the Si-Si percolation splitting. In the experimental difference spectrum, such shift may lead to an artificial screening of the antiresonance due to the minor Si-Si component, as predicted within the percolation scheme when assuming a strict invariance of the Raman frequencies.

For unambiguous insight into the particular $k$-dependence of the Si-Si doublet, we may thus need to consider a larger $k$-difference. Due to the lack of experimental data, we opt for a direct *ab initio* insight.

**IV. Percolation vs. *ab initio* insights – (Si-Si) doublet**

*Ab initio* SiGe Raman spectra are calculated by using a pseudopotential spin density-functional supercell code (AIMPRO),[29,30] along the local exchange-correlation parametrization by Perdew and Wang,[31] taking the potentials for Si and Ge as proposed by Hartwigsen *et al.* (Ref. **32**). In doing so, we follow the same procedure earlier adopted to optimize SiGe supercells (Ref. **33**), but generally using a thinner 12×12×6 $k$-point mesh as proposed by Monkhorst and Pack (Ref. **34**). A representative *ab initio* insight is obtained by using large (32-atom) tetragonal Si$_{0.5}$Ge$_{0.5}$ supercells of three different types, corresponding to local clustering ($k$>0) and local anticlustering ($k$<0) besides random Si↔Ge substitution ($k$=0). The finite $k$ values are taken symmetrical and as large as possible while maintaining at the same time a minimum bond counting of ten per species (out of 64 bonds in total per supercell) for a reliable statistics, being clear that the statistics of the alloy disorder is anyway not fully taken into account due to the finite size of the supercell. A relevant set of $k$ values matching the above criteria is (+0.31, 0, -0.31), corresponding to the (Ge-Ge, Si-Ge, Si-Si) bond counting per supercell of (21, 22, 21), (16, 32, 16) and (11, 42, 11), respectively, as governed by Eqs. (10). The Si (small-blue symbol) and Ge (large-yellow symbol) atom arrangements in the successive (111) planes of the supercells in question are shown in **Fig. 4**.

Interestingly, such $k$-set covers three distinct types of intensity-interplays within the Si-Si percolation doublet when $k$ varies, in reference to **Fig. 3**. One in which the lower feature is just absent ($k$~0.3), one in which the intensity ratio between the lower and upper features reaches one third ($k$=0), and one in which the latter intensity ratio is inverted ($k$~-0.3). This is ideal to test the percolation scheme from *ab initio* calculations. In principle, the refined percolation trend related to the fine structure of the Si-Si doublet, as indicated above, should come on top of the basic MREI-like trend reflecting the relative abundance of the three bond species in the crystal when $k$ varies. As already mentioned, in case of clustering this corresponds to reinforcement of the Raman signals due to the 'homo' Ge-Ge and Si-Si bonds at the cost of the Raman signal due to the hetero Si-Ge bond, and *vice versa* in case of anticlustering.

The *ab initio* Raman spectra obtained with the $k$=(+0.31, 0, -0.31) supercells shown in **Fig. 4** are displayed in (top, medium, bottom) positions of **Fig. 5** (thick curves), correspondingly. A line broadening of 4 cm$^{-1}$ is taken, i.e. large enough to be realistic, but small enough to allow the resolution of neighboring percolation features (the minimum splitting between neighboring percolation features is ~10 cm$^{-1}$, see **Fig. 1**). An additional series of Raman spectra obtained by inverting the Si and Ge atoms in each supercell, thereby leaving the $k$ value unchanged, is slightly shifted beneath the original series, for comparison (thin curves). The two series are useful to distinguish intrinsic trends, i.e. visible in the two *ab initio* Raman spectra corresponding to a given $k$ value, that deserve attention, from merely fortuitous ones, i.e. due to a particular arrangement of the Si and Ge atoms, that should not be discussed.

1. **Basic assignment via limit *ab initio* frequencies**

A prerequisite before proper comparison of the *ab initio* Raman spectra and of the corresponding percolation-type Raman lineshapes, respectively calculated within a supercell approach (3D) and along the LCA (1D), is to bridge the 3D – 1D gap in the discussion of the *ab initio* data. This comes to assign each *ab initio* Raman line in terms of a given bond-stretching. In principle this can be done by examining the individual *ab initio* vibration pattern behind each Raman line. However, in practice, this is not feasible, due to the complexity of the lattice relaxation in our disordered Si$_{0.5}$Ge$_{0.5}$ supercells, which distorts and finally obscures the atom displacements. Based on our experience,[3,5,6,10,12] such direct assignment is possible only when using pure supercells, or very basic impurity motifs diluted in an otherwise pure host matrix.

We thus proceed along the latter line and delimit the Ge-Ge, Si-Si and Si-Ge *ab initio* spectral ranges by using a reduced set of Raman frequencies obtained by considering ideally simple impurity arrangements taken as representative of the corresponding limit bond-stretchings when approaching full c/ac. In case of clustering, such limits frequencies are accessed by using pure Ge (oblique-right hatching, see **Fig. 5**) and Si (oblique-left hatching) supercells, for the 'homo' Ge-Ge and Si-Si bonds (1 mode per bond), and isolated Ge (grey) or Si (white) impurities in otherwise pure Si or Ge supercells, respectively, for the 'hetero' Si-Ge bond (2 modes in total). The corresponding modes, denoted 'Ge' (x=1; $k$=1; 304 cm$^{-1}$), 'Si' (x=0; $k$=1; 522 cm$^{-1}$), 'Si:Ge' (x~0; $k$~1; 451 cm$^{-1}$) and 'Ge:Si' (x~1; $k$~1; 385 cm$^{-1}$), respectively, as observable also in **Fig. 1**, are marked by arrows/open-red squares at the top of **Fig. 5**. In case of full anticlustering the starting supercell is a zincblende SiGe one (crossed hatching), formed with two intercalated Ge and Si *fcc* sublattices, thus providing a natural insight into the 'pure Si-Ge' frequency. The alternative Ge-Ge and Si-Si impurity frequencies are accessed by realizing a unique Si↔Ge substitution either on the Si sublattice or on the Ge one, respectively. The corresponding modes, denoted as 'SiGe' (x=0.5; $k$=-1, ~421 cm$^{-1}$) for the pure zincblende SiGe crystal, and 'SiGe:Ge' (x~0.5, $k$~-1, ~272 cm$^{-1}$) or 'SiGe:Si' (x~0.5, $k$~-1, ~473 cm$^{-1}$), for the Ge and Si impurity modes in zincblende SiGe, respectively, are marked by arrows/plain-blue squares at the bottom of **Fig. 5**. They are correspondingly indicated in **Fig. 1**.

Most of the above-quoted *ab initio* frequencies were already identified in our previous work on SiGe,[6] except the 'SiGe:Ge' and 'SiGe:Si' frequencies which have required a novel *ab initio* insight in this work. The isolated Ge and Si atoms on the foreign Si-like and Ge-like *fcc* sublattices of the zincblende SiGe supercell generate distinct impurity modes on each side of the main Raman mode due to the host crystal. Due to the $T_d$ symmetry of the site occupied by such impurities, we expect triply-degenerated impurity modes. However, this is not so in practice when using 32-atom supercells. Due to the moderate size of the 32-atom supercells, each isolated impurity in one supercell "feels" the influence of the next impurities in the surrounding supercells. The influence is not same depending on whether the next impurities in question belong to the plane formed by the short axes of the reference tetragonal supercell, or are situated along its long axis (see **Fig. 4**). Altogether this produces a lowering of the impurity-site symmetry, leading to the emergence of a distinct doublet per impurity (corresponding to a significant frequency gap of ~10 cm$^{-1}$, not shown), instead of the expected singlet. Such drawback was overcome by using larger 64-atom supercells, as already done for a similar purpose in our previous SiGe work (Ref. **6**). In this case, as expected, each impurity provides a unique Raman line, located at the above-quoted 'SiGe:Ge' or 'SiGe:Si' frequency. The corresponding Raman spectra, obtained by using a 6×6×6 $k$-point mesh and a typical line broadening of 4 cm$^{-1}$, are shown in **Fig. 6**.

We can be more explicit and assign the limit (x~0,1) *ab initio* Raman modes within the [1x(Ge-Ge), 3x(Si-Ge), 2x(Si-Si)] percolation scheme. 'Ge' and 'SiGe:Ge' both refer to the same unique (Ge-Ge) percolation branch, that covers all possible Ge-Ge stretching modes in the SiGe alloy. 'Ge:Si', 'Si:Ge' and 'SiGe' refer to Si-Ge stretching in a pure Ge environment, in a pure Si one and in a (Si,Ge)-mixed one, respectively, thus identifying with the side $(Si-Ge)_2^{Ge}$ and $(Si-Ge)_2^{Si}$ percolation branches and with the intermediary $(Si-Ge)_2^{SiGe}$ one, respectively. As for 'SiGe:Si' and 'Si' they are due to Si-Si stretching in a pure-Ge environment and in a pure-Si one, respectively, thus assimilating with $(Si-Si)_1^{Ge}$ and $(Si-Si)_1^{Si+SiGe}$, respectively. In summary, our reduced set of seven *ab initio* frequencies covers the whole set of percolation oscillators, none is left by. The Ge-Ge, Si-Ge and Si-Si spectral ranges are accordingly delimited using dotted rectangles in **Fig. 5**.

### 2. Raman – intensity aspect

Now, we discuss the intensities of the as-identified *ab initio* Raman signals in their $k$-dependence. The general trend is that the Raman signals due to the 'homo' Ge-Ge and Si-Si bonds progressively reinforce with clustering, at the cost of the intermediary Raman signal due to the 'hetero' Si-Ge bond, as expected. This is consistent both with the available experimental data, in reference to point **(i)** (see **Sec. II**), and also with the predictions of the MREI and percolation schemes (see **Fig. 3**). We want to be more accurate and examine further the $k$-dependence of the Si-Si and Si-Ge *ab initio* fine structures (being clear that the MREI scheme falls short of explaining any fine structure related to any bond).

The *ab initio* insight is no so helpful with respect to the Si-Ge signal, in that no one-to-one correspondence can be established between the Si-Ge percolation triplet and the corresponding *ab initio* Raman pattern. In particular, the minor/upper component of the Si-Ge percolation triplet ideally represented by a unique feature in **Fig. 3**, appears to decompose into a myriad of *ab initio* lines already in the random alloy ($k$=0) – refer to the stars in **Fig. 5**. Such shortcoming of the SiGe percolation scheme far-off the Si-rich limit has already been pointed out in **Sec. I** and also in Ref. **6**. A promising trend, nevertheless, is that, in comparison with the main/central Si-Ge feature, such extended fine structure is less represented at $k$=-0.31, in spite of the large number of Si-Ge bonds, than at $k$=+0.31, in spite of the small number of Si-Ge bonds. This seems to indicate a relative reinforcement with clustering. A similar trend may be inferred for the minor/lower component of the Si-Ge triplet from the pronounced asymmetry of the main/central Si-Ge mode on its low-energy side at $k$=+0.31. Altogether, this is consistent with the percolation scenario (see **Fig. 3** and **Sec. III-2**), and also with the existing experimental data, in reference to point **(ii)** (see **Sec. II**). However, further *ab initio* calculations, involving much larger SiGe supercells, in hope that the enlarged statistics on the Si-Ge bonds leads to a simplification of the extended Si-Ge fine structure (marked by stars in **Fig. 5**) into well-defined – possibly broadened – Raman features, are definitely needed in view to develop a meaningful comparison with the percolation triplet. Such extended *ab initio* calculations fall beyond the scope of this work. Our present policy with respect to the Si-Ge signal was rather to rely directly on experimental data, as detailed in **Sec. III-3**.

The *ab initio* insight is much more fruitful regarding the Si-Si fine structure, our main concern in this **Sec**. The *ab initio* Si-Si signal at $k$=0 decomposes into a distinct lower-minor/upper-dominant doublet separated by ~10 cm$^{-1}$ (refer to the modes numbered 1 and 2 in **Fig. 5**) which strongly resembles its percolation analogue (refer to the central curve in **Fig. 3**). Moreover, only the upper mode survives at $k$=+0.31, while the lower mode has turned dominant at $k$=-0.31. Altogether, this is consistent with the available experimental data, in reference to **(iii)** (see **Sec. II**) Most of all, this conforms ideally to the percolation scenario, as apparent in **Fig. 3**. This is just the kind of independent *ab initio* support we needed, i.e. as concerned with the impact of local c/ac at the ultimate scale of the fine structure of the Raman signal due to a given bond, i.e. Si-Si in this case, to validate our generalized $k$-dependent version of the SiGe percolation scheme.

### 3. Raman – frequency aspect

Last, we cannot escape a brief discussion of the $k$-dependence of the Si$_{0.5}$Ge$_{0.5}$ Raman frequencies in the *ab initio* spectra, though this is not of direct use in this work centered on the Raman intensities. The effects of clustering ($k$=+0.31) and anticlustering ($k$=-0.31) are separately considered below, in reference to the random case ($k$=0). Accordingly the *ab initio* frequencies of the unique Ge-Ge, main Si-Ge and Si-Si doublet in the latter case are emphasized in **Fig. 5** (refer to the plain-blue circles at $k$=0), and in **Fig. 1** as well. Generally, in the discussion below our attention is focused on the $k$-dependence of the dominant feature in each spectral range (see **Fig. 3**), which fix the trend.

With clustering ($k$>0), the unique Ge-Ge mode massively blue-shifts. In contrast the main/central Si-Ge mode undergoes a clear red-shift. As for the upper/dominant Si-Si mode, only a slight blue-shift can be detected. Such *ab initio* shifts, are globally consistent in nature, if not always in magnitude, with the existing experimental data in the literature. Unambiguous trends concerning the first two modes were evidenced by Schorer *et al*. in Ref. **19** (see Fig. 3 therein) in their careful step-by-step Raman study of an as-grown Si$_8$Ge$_8$ superlattice submitted to progressive annealing until complete destruction of the long range order, i.e. achievement of a random Si↔Ge substitution. However, such study is not so helpful with respect to the remaining Si-Si signal. This is because the Si-Si signal develops into a proper alloy-related mode only in the very final stage of annealing, corresponding, in fact, to the last-recorded Raman spectrum of Schorer *et al*. Nevertheless we can appreciate in the alternative data of Schorer *et al*. reported together with those of Tsang *et al*. in **Fig. 2**, that clustering induces a slight blue-shift of the upper/dominant Si-Si mode (refer to the double-horizontal arrow in **Fig. 2b**). Recently, a detailed study of the Raman shifts of the main Ge-Ge, Si-Ge and Si-Si modes of a Si$_{0.53}$Ge$_{0.47}$ epitaxial layer, performed by Reparaz *et al*.,[27] has revealed massive blue-shifts for all Raman modes under annealing. These were, at least partially, attributed to a progressive suppression, under annealing, of compositional-inhomogeneity due to Ge-clustering in the as-grown sample. Apparently this contradicts the present *ab initio* trend, at least for the main/central Si-Ge mode. However, the experimental shifts observed by Reparaz *et al*. may be dominantly due to a progressive relaxation of the internal and substrate-induced strains in the as-grown epilayers, as also envisaged by the authors. Indeed no variation in the relative Raman intensities of the main Ge-Ge, Si-Ge and Si-Si features, the sign of c/ac referring to Eqs. (10), can be detected in the data reported by Reparaz *et al* (refer to Fig. 1b of Ref. **27**), even at an attentive sight.

Such *ab initio* shifts of the unique Ge-Ge (blue-shift, large), main/central Si-Ge (red-shift, large) and upper/dominant Si-Si (blue-shift, small) modes find a natural explanation within the percolation scheme. Referring to the frequency-map reported in **Fig. 1**, it is just a matter to progress along the corresponding percolation branches from the random alloy (x=0.5, $k$=0, plain-blue circles) towards the relevant pure-Si or/and pure-Ge crystal(s) – the final products in case of full clustering (x=0 and 1, $k$=+1, open-red squares), as indicated by the straight (Ge-Ge, upper Si-Si and side Si-Ge modes) or curved (main Si-Ge mode) red-arrows. For example, with clustering the Ge-Ge bonds tend to vibrate more in their own Ge-like environment, thus ultimately converging to the vibration of the pure Ge crystal (x=1, $k$=1, open-red square). Similarly, the upper-dominant Si-Si mode, due to Si-Si vibration in its own Si-like environment, should ultimately look like the vibration of the pure Si crystal (x=0, $k$=1, open-red square). As for the main Si-Ge mode, due to Si-Ge stretching in a (Si,Ge)-mixed environment, this may ultimately arise as well from the Si-like or Ge-like parts of the phase-separated alloy, as an impurity mode in both cases then. Accordingly, the progression is double, as reflected by antagonist curved arrows simultaneously converging onto the vibration frequencies of Si or Ge pairs of impurities in next-nearest neighbor positions (the basic motif corresponding to Si-Ge stretching in a mixed environment as extensively explained in Ref. **6**) immersed in otherwise pure Ge (x~1, $k$=1, not shown) or Si (x~0; $k$=1, not shown) crystals, respectively.[6] Eventually, the two progressions add to generate an overall Si-Ge red-shift. Note that the central Si-Ge mode collapses for $k$ ≥0.5 (see **Fig. 3**), being relayed by its lower and upper satellites. These refer to Si-Ge stretching in pure-Ge and pure-Si environments, respectively, ultimately assimilating with the Ge:Si (x~1, $k$=1, open-red square) and Si:Ge (x~0, $k$=1, open-red square) impurity modes, as indicated by the opposite dotted red-arrows (plain red-arrows are kept for the dominant modes in the considered *ab initio* $k$-range in **Fig. 5**, for more clarity).

With anticlustering ($k$<0), we observe in **Fig. 5** that the *ab initio* frequencies of all dominant modes remain quasi invariant. The reference in this case is the pure SiGe zincblende crystal (x=0.5, $k$=-1), i.e. the ultimate host crystal in case of full anticlustering. The phonon shifts are thus discussed from **Fig. 1** by using curved (semi-closed loops) blue-arrows starting from frequencies identified on the relevant percolation branches of the random SiGe alloy (x=0.5, $k$=0, plain-blue dots in **Figs. 1** and **5**) and ending at the corresponding frequencies in zincblende SiGe (x=0.5, $k$~-1, plain-blue squares in **Figs. 1** and **5**). For example, the central/main Si-Ge mode, due to Si-Ge vibration in a (Si,Ge)-mixed environment naturally converges onto the frequency of the pure zincblende SiGe (x=0.5, $k$=-1, plain-blue dot, centre of **Fig. 1**). As for the unique Ge-Ge mode and lower/main Si-Si component of the Si-Si doublet, these assimilate *in fine* with the SiGe:Ge (x=0.5, $k$~-1, plain-blue dot, bottom of **Fig. 1**) and SiGe:Si (x=0.5, $k$~-1, plain-blue dot, top of **Fig. 1**) impurity modes. We recall that the latter correspond to a unique Ge↔Si substitution on the *fcc* Si or Ge sub-lattices of zincblende SiGe, respectively. In each case, the starting and ending frequencies differ by less than a few centimeters inverse, corresponding in practice to quasi invariance of the Raman frequencies.

**V. Conclusion**

We report on a comprehensive study of the intensities of the Raman multi-lines of the diamond-type SiGe alloy in their (co-)dependence on local clustering/anticlustering within the *percolation* scheme, which has recently lead to a novel understanding of the random-SiGe Raman pattern along the linear chain approximation in terms of six basic one-dimensional (1D) oscillators [1×(Ge-Ge), 3×(Si-Ge), 2×(Si-Si)].[6] For doing so, the individual fractions of such 1D-oscillators are expressed via a relevant order parameter $k$. This is introduced along the approach used by Verleur and Barker in their 1D *cluster* model that distinguishes between all possible first neighbor oscillators of a three-dimension (3D) zincblende alloy.[8] However, care is taken to overcome some ambiguity behind the 1D – 3D ambivalence in the treatment of Verleur and Barker by adopting a pure 1D formalism in the present SiGe version. Moreover the formalism is generalized to the SiGe-like diamond structure. Also, not only the first-neighbor environments of a bond are covered, but also the second-neighbor ones, in reference to the Si-Si doublet and to the Si-Ge triplet, respectively. Last, the SiGe-diamond version takes into account inter-bond transfer of oscillator strength on top of the zincblende-like intra-bond transfer of oscillator strength driven by local clustering/anticlustering. Predictive $k$-dependent Si$_{0.5}$Ge$_{0.5}$ percolation Raman lineshapes compare fairly well with (i) existing experimental data, when referring to the Si-Ge fine structure, and with (ii) *ab initio* Raman spectra calculated by using large (32-atom) disordered supercells matching the required $k$ values, referring then to the Si-Si fine structure.

Altogether, the present work formalizes an intrinsic ability behind Raman scattering to achieve a quantitative insight into clustering/anticlustering in an alloy at the very local scale. This was still lacking so far. Interestingly, such formalization is achieved along the linear chain approximation, i.e. simply at 1D.

**Acknowledgements**

The authors are grateful to A. V. Postnikov for a critical reading of the manuscript and many useful discussions, and also to M. I. Alonso for stimulating our interest on clustering/anticlustering issues in SiGe. This work was supported by the European funding of Region Lorraine under Project FEDER-Percalloy n° presage 34619.

**Figure captions**

**Fig. 1:** (color online) Schematic [1x(Ge-Ge), 3x(Si-Ge), 2x(Si-Si)] percolation scheme for random $Si_{1-x}Ge_x$, as directly inspired from Ref. **6** (refer to Figs. 1 and 4 therein). On the right, the individual oscillators are labeled according to their standard percolation terminology, i.e. using a main term equipped with a superscript and a subscript, in reference to the considered bond-stretching and to the environment in which it takes place, with respect to both composition and length scale, respectively. A numerical labeling is also used (1 – 6), on the left, for more convenience, notably in a comparison with **Figs. 3** and **5**. Limit *ab initio* frequencies calculated in diamond-Si (x~0, oblique-left hatching) and diamond-Ge (x~1, oblique-right hatching), on the one hand (open-red squares), and in zincblende-SiGe (x~0.5, crossed hatching), on the other hand (plain-blue squares), using either pure supercells or containing a unique impurity (as schematically indicated), are used for a qualitative discussion of the frequency shifts of the main $Si_{0.5}Ge_{0.5}$ Raman features induced by clustering (red straight-curved arrows) or anticlustering (blue semi-closed loops/arrows). The *ab initio* frequencies of the unique Ge-Ge, main Si-Ge and Si-Si doublet in random-$Si_{0.5}Ge_{0.5}$, taken from the central curves ($k$=0) in **Fig. 5**, are added (plain-blue circles), for reference purpose. Globally, the same schematic code and labeling of the limit Raman frequencies is used in **Fig. 5**. The $k$-dependence of the individual fractions of oscillators, which monitor directly the Raman intensities, is expressed via the $P_{SiSi}$ and $P_{GeGe}$ probabilities (see text) in the body of the figure.

**Fig. 2:** (color online) Representative $Si_{0.5}Ge_{0.5}$ Raman spectra taken from the literature, used to reveal the effect of local clustering on the (a) Si-Ge (data digitalized from Fig. 1 of Ref. **16**) and (b) Si-Si (data digitalized from Fig. 1 of Ref. **20**) fine structures, as emphasized by thick arrows in each panel. The spectra refer to epitaxial layers grown as random alloys (bottom curves in each panel) or under the form of superlattices (upper curves in each panel), corresponding either to a moderate clustering [$\eta$=0.64, panel (a)] or to a $Si_4Ge_4$ sequence with interface mixing [panel (b)]. The stars refer to the underlying Si substrate in each case. In panel (a) the upper spectrum is the difference Raman spectrum obtained by subtracting the Raman spectra of the random alloy (bottom curve) from that of the superlattice ($\eta$=0.64), multiplied by 4 (as indicated). Corresponding percolation-type lineshapes (thick-red curves) for the random (bottom curve, $k$=0) and clustered (upper curve, $k$=0.2) $Si_{0.5}Ge_{0.5}$ alloys are superimposed to the experimental data (thin-black ones), for comparison. The horizontal double-arrows mark significant phonon shifts with clustering.

**Fig. 3:** (color online) $k$-dependent percolation (black-thick curves) and MREI (black-thin curves) $Si_{0.5}Ge_{0.5}$ Raman lineshapes in case of clustering ($k$>0) and anticlustering ($k$<0), calculated by using the individual $k$-dependent fractions of oscillators given in **Fig. 1**. The Raman frequencies and phonon damping are taken constant, identical to those in the random $Si_{0.5}Ge_{0.5}$ alloy, in a crude approximation (see text, and Ref. **6**). The percolation-type Raman spectra corresponding to the selected $k$ values of 0.31, 0 and -0.31 are emphasized (red curves). A direct comparison can be made with corresponding $k$-dependent *ab initio* Raman spectra reported in **Fig. 5**. Special attention may be awarded to the sensitive Si-Si doublet identified by a specific labeling.

**Fig. 4:** (color online) Positioning of the Si (small-green symbol) and Ge (large-yellow symbol) atoms in three selected 32-atom $Si_{0.5}Ge_{0.5}$ supercells corresponding to clustering ($k$=+0.31, left position) random substitution ($k$=0, center position), and anticlustering ($k$=-0.31, right position). Identical $k$ values are obtained just by inverting the Si and Ge atoms in each supercell.

**Fig. 5:** (color online) *Ab initio* Raman spectra obtained with the three 32-atom Si$_{0.5}$Ge$_{0.5}$ supercells displayed in **Fig. 4** (thick curves), corresponding to local clustering ($k$=+0.31, top position), random substitution ($k$=0, medium position) and local anticlustering ($k$=-0.31, bottom position). Additional *ab initio* Raman spectra obtained by inverting the positions of the Si and Ge atoms in each supercell (thin curves), thereby leaving the $k$ values unchanged, are shifted beneath the original curves, for comparison and identification of intrinsic trends. The *ab initio* frequencies of the unique Ge-Ge, main Si-Ge and Si-Si doublet in random-Si$_{0.5}$Ge$_{0.5}$ are pointed out (plain-blue circles), for reference purpose. The Ge-Ge, Si-Ge and Si-Si spectral ranges, delimited by dotted rectangles for help in the discussion, are identified based on limit *ab initio* frequencies when approaching full clustering (top-red arrows, open-red squares) and full anticlustering (bottom-blue arrows, plain-blue squares). Globally, the same schematic code and labeling of such limit Raman frequencies is used as in **Fig. 1**. The sensitive Si-Si doublet is emphasized by using numbers (1,2), for unambiguous comparison with the corresponding percolation-type Raman features in **Fig. 3**.

**Fig. 6:** (color online) *Ab initio* Raman spectra obtained with two 64-atom zincblende SiGe supercells containing either one isolated Si impurity on the *fcc* Ge sublattice (top spectrum) or one isolated Ge impurity on the *fcc* Si sublattice (bottom spectrum), as schematically indicated. Distinct modes due to the SiGe-zincblende host matrix and to the isolated Ge and Si impurities are labeled using the same symbol/color code as in **Figs. 1** and **5**, for a direct correspondence.

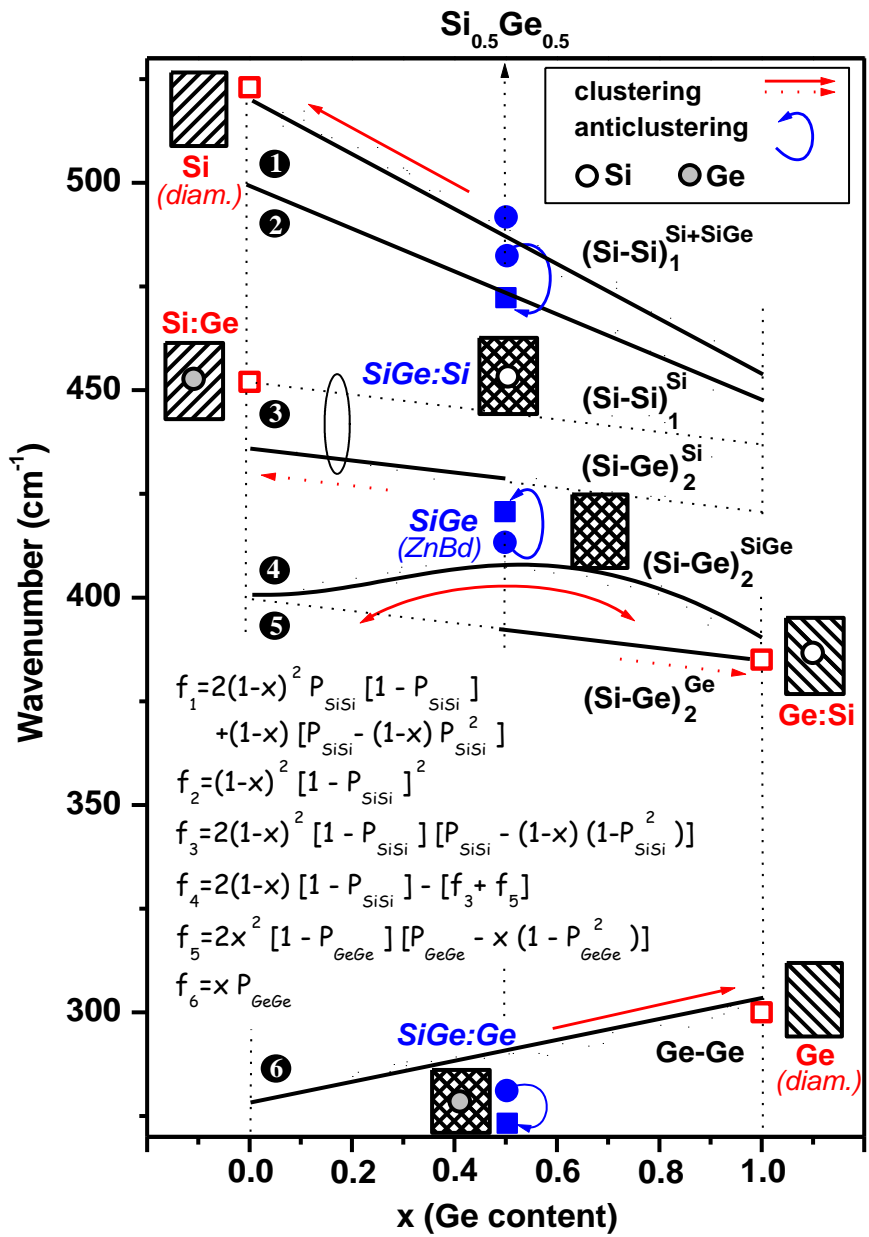

**Figure 1**

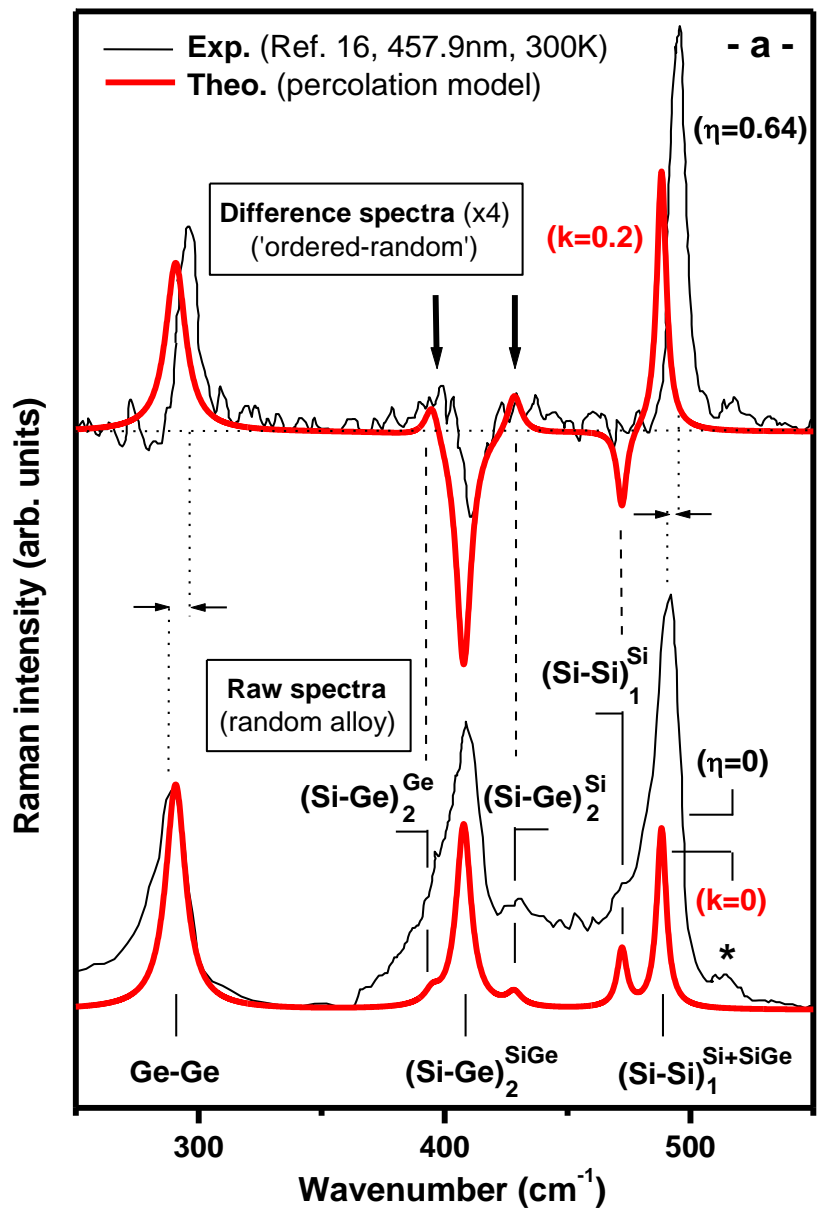

**Figure 2a**

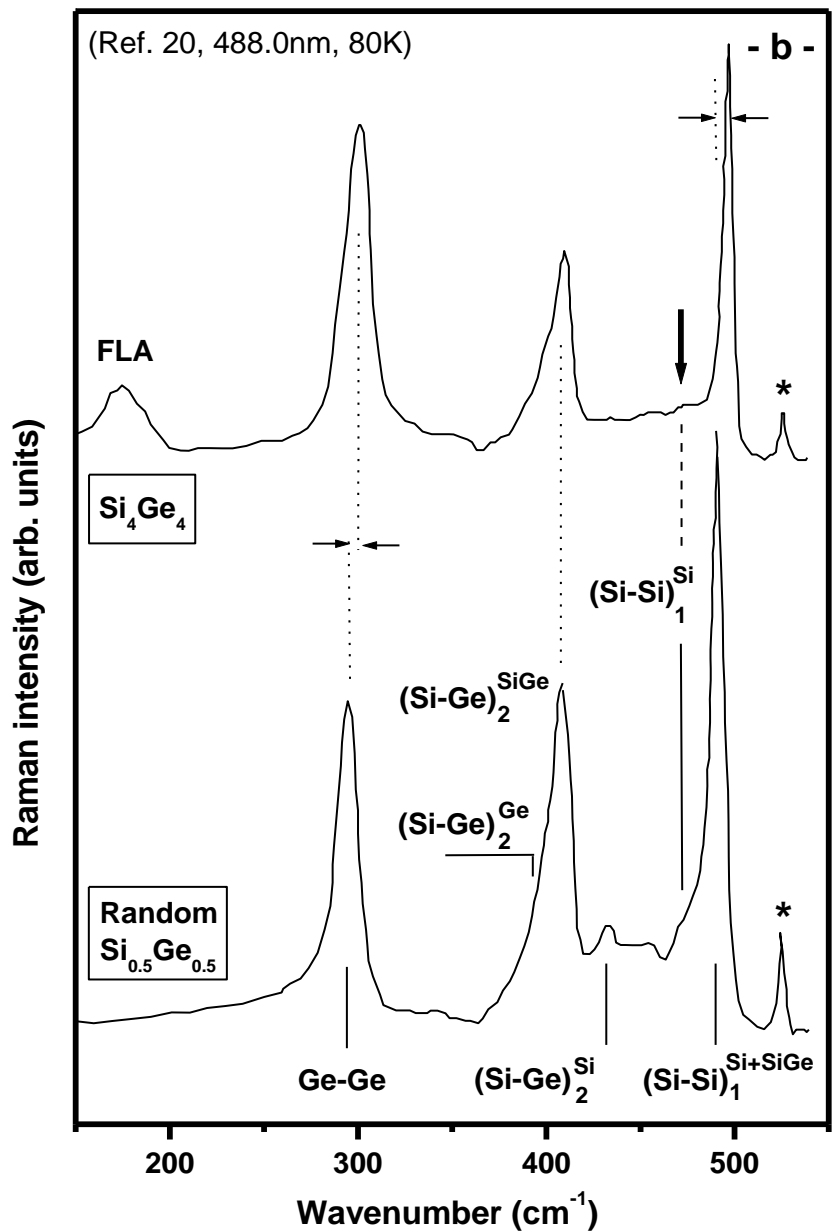

Figure 2b

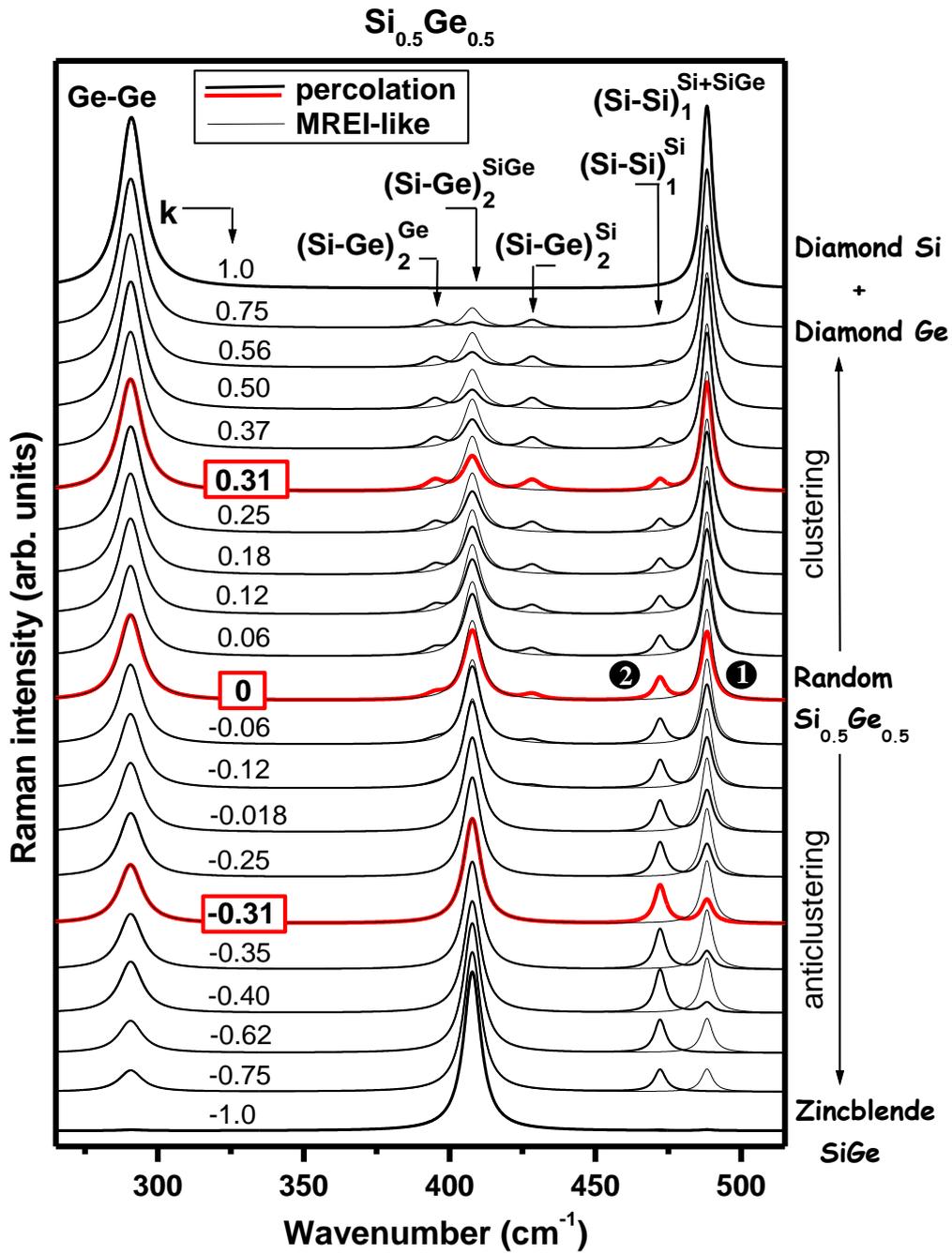

Figure 3

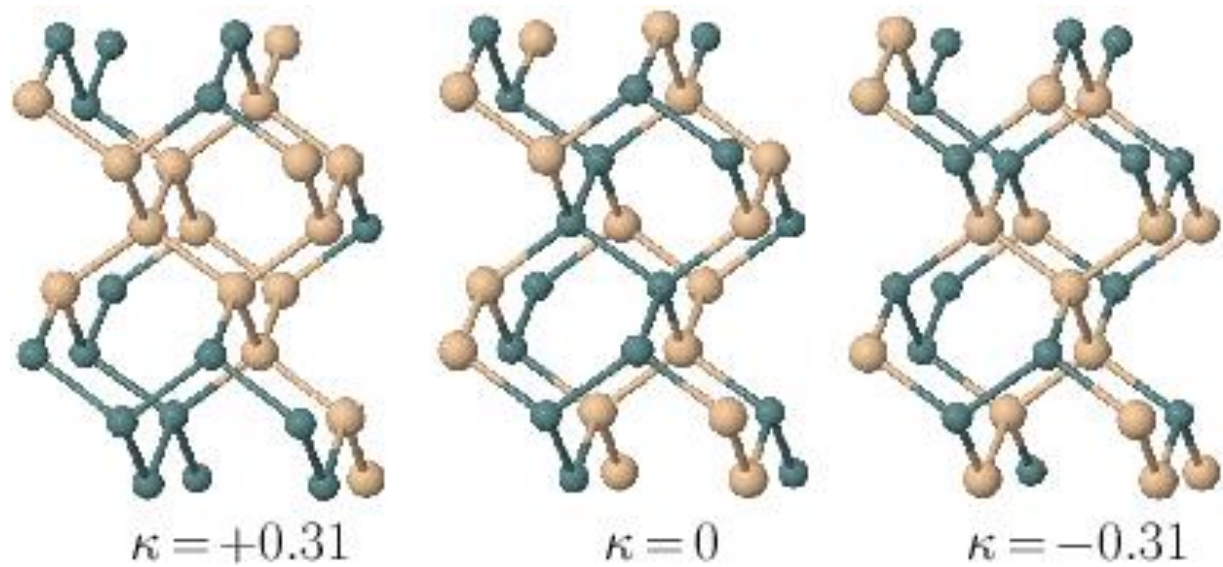

**Figure 4**

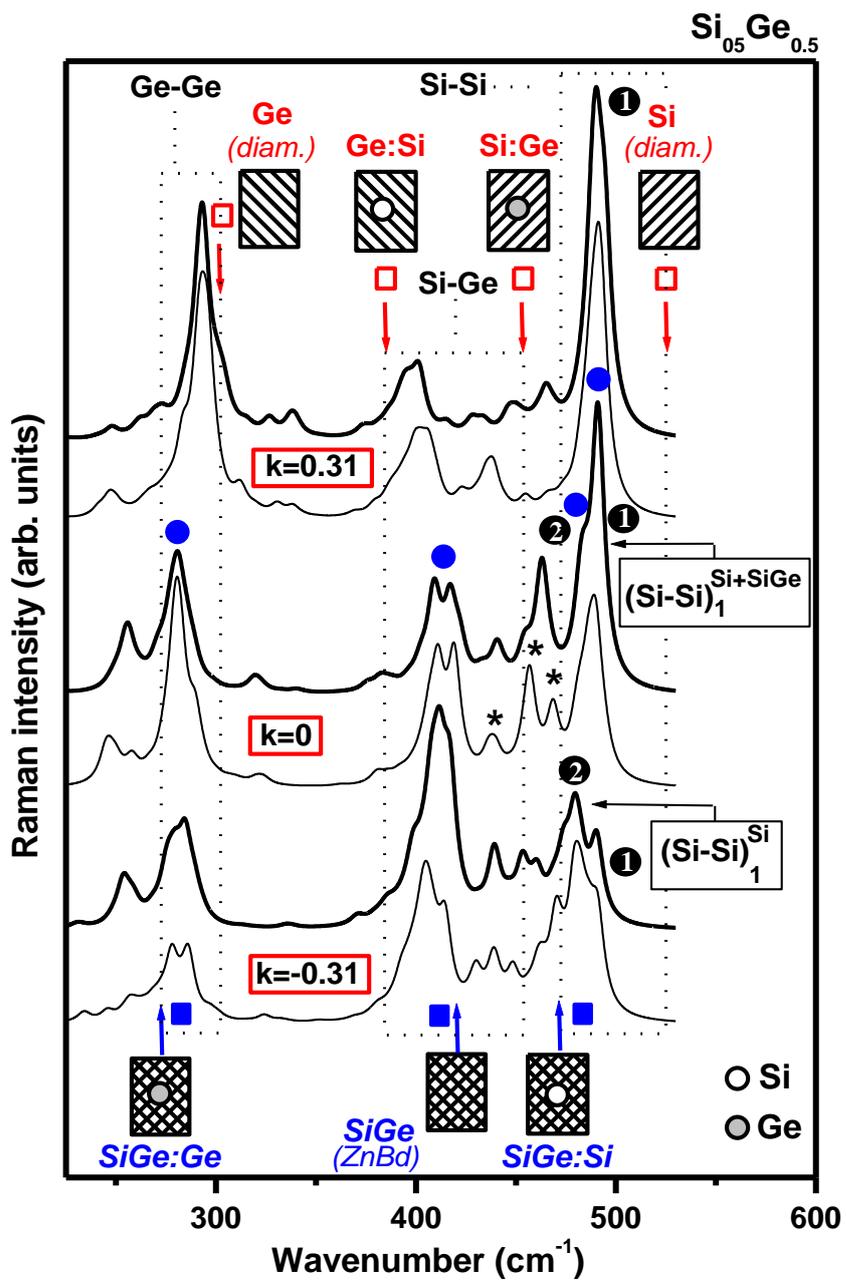

Figure 5

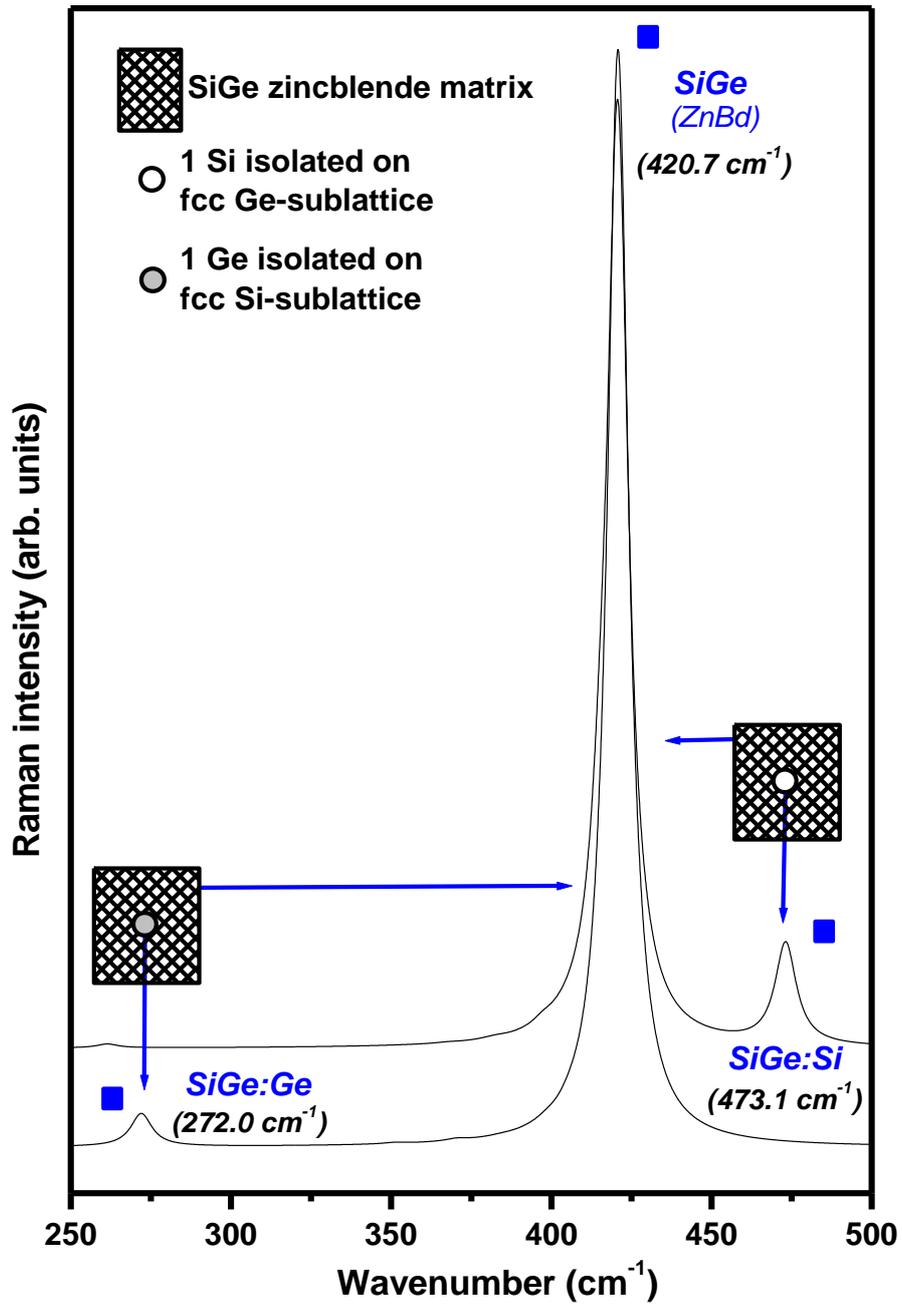